\documentclass[twocolumn]{aastex62}
\usepackage{ifpdf} 
\newif\ifpdf
\ifx\pdfoutput\undefined
     \pdffalse 
\else
\pdfoutput=1 
\pdftrue
\fi

\usepackage{amsmath}
\usepackage{amsfonts}
\graphicspath{{./}{figures/}}
\submitjournal{ApJ}
\shorttitle{IBS Synchrotron Emission}
\shortauthors{Kandel et al.}
\defcitealias{2016ApJ...828....7R}{RS16} 	
\begin{document}

\title{The Synchrotron Emission Pattern of IntraBinary Shocks}

\correspondingauthor{R.W. Romani}
\email{rwr@stanford.edu}

\author[0000-0002-5402-3107]{D. Kandel}
\affil{Department  of  Physics,  Stanford  University,  Stanford,  CA, 94305, USA}

\author{Roger W. Romani}
\affil{Department  of  Physics,  Stanford  University,  Stanford,  CA, 94305, USA}

\author{Hongjun An}
\affil{Department of Astronomy and Space Science, Chungbuk National University, Cheongju, 28644, Republic of Korea}

\begin{abstract}
We model millisecond pulsars winds colliding with radiatively-driven companion winds in black widow and redback systems. For the redbacks, the geometry of this intrabinary shock (IBS) is quite sensitive to the expected equatorial concentration in the pulsar outflow. We thus analytically extend IBS thin-shock models to $\sim \sin^{2n}\theta$ pulsar winds. We compute the synchrotron emission from such shocks, including the build-up and cooling of the particle population as it accelerates along the IBS. For reasonable parameters, this IBS flux dominates the binary emission in the X-ray band. The modeling shows subtle variation in spectrum across the IBS peak, accessible to sensitive X-ray studies. As example applications, we fit archival {\it CXO}/{\it XMM} data from the black widow pulsar J1959+2048 and the redback PSR J2339-0533, finding that the model reproduces well the orbital light curve profiles and energy spectra. The results show a very hard injected electron spectrum, indicating likely dominance by reconnection. The light curve fitting is sensitive to the geometric parameters, including the very important orbital inclination $i$. Coupled with optical fits of the companion star, such IBS X-ray light curve modeling can strongly constrain the binary geometry and the energetics of the MSP wind.

\end{abstract}

\keywords{gamma rays: stars - pulsars: individual (J1959+2048, J2339-0533)}

\section{Introduction} \label{sec:intro}

In the so-called 'spider' binaries, millisecond pulsar heating drives winds from the non-degenerate companions which occlude the pulsed radio signal; accordingly, few such objects were found in classical radio pulsar surveys. However, the penetrating $\gamma$-ray emission from such pulsars makes them prominent in the {\it Fermi} LAT sky and dedicated searches of LAT unidentified sources have turned up many such objects, leading to a renaissance in the study of this intense interaction phase. The objects are generally divided into `redbacks' (hereafter RB) with $M_c \approx 0.1-0.4 M_\odot$ stellar mass companions and `black widows' (BW) orbited by $M_c \le 0.04 M_\odot$ sub-stellar objects. One may identify sub-classes of the RB: `Transitioning MSP' (Tr) when the companion's Roche-lobe overflow drives it to an intermittent accretion phase and `Huntsman' (Hu) MSP when the companion is in the giant phase. For the BW, a `Tidarren' (Ti) subclass identifies those exceptionally short $P_B$, low $M_c$ systems where the companion is hydrogen-stripped; these extend down to short-period pulsar-planet MSP.
    
While the GeV $\gamma$-ray signal dominates the total photon output of these systems, in the optical, the companion, often with a strongly heated face, dominates. In the X-rays, emission is seen from the pulsar magnetosphere and the heated companion surface, but for many systems, especially for RB, the orbital light curve presents strong peaks of hard-spectrum $\Gamma \approx 1-2$ emission that appears to arise from an intrabinary shock (IBS) between the relativistic pulsar wind and a massive wind driven from the companion. Here the stellar wind/pulsar momentum flux ratio 
\begin{equation}\label{eq:beta}
\beta = {\dot M_w} v_w c/{\dot E}_{\mathrm{PSR}}
\end{equation}
controls the shock geometry \citep[hereafter RS16]{2016ApJ...828....7R}. In this paper we further explore the synchrotron spectrum of this IBS component.
\bigskip
\bigskip
\section{Anisotropic pulsar wind -- Stellar wind Interaction}\label{sec:shock}

Numerical simulations of pulsar magnetospheres \citep[e.g.][]{2016MNRAS.457.3384T}
suggest that the energy/momentum flux of the pulsar wind is equatorially concentrated as $\propto \sin^n\theta_*$ with $n=2$ or even $n=4$ for an oblique rotator ($\theta_*$ is the angle from the pulsar spin axis, assumed perpendicular to the orbital plane). In the \citetalias{2016ApJ...828....7R} implementation of IBS shock emission in the {\tt ICARUS} binary light curve modeling code, we allowed the pulsar power to be distributed $\propto \sin^n\theta_*$, but this power illuminates an IBS shock computed from the contact discontinuity (CD) shape for the collision of two cold isotropic winds described by \cite{1996ApJ...469..729C}. This is a good approximation when the pulsar wind dominates ($\beta \ll 1$, the `black widow' BW case), but there are significant shape distortions when $\beta>1$ and the shock wraps around the pulsar (the RB case). Here, we provide expressions for the shape for both the cases, for a thin shock (efficient cooling) with the pulsar wind scaling as $\sin^{2n} {\theta}$.

\begin{figure}
\centering
\includegraphics[scale=1.0]{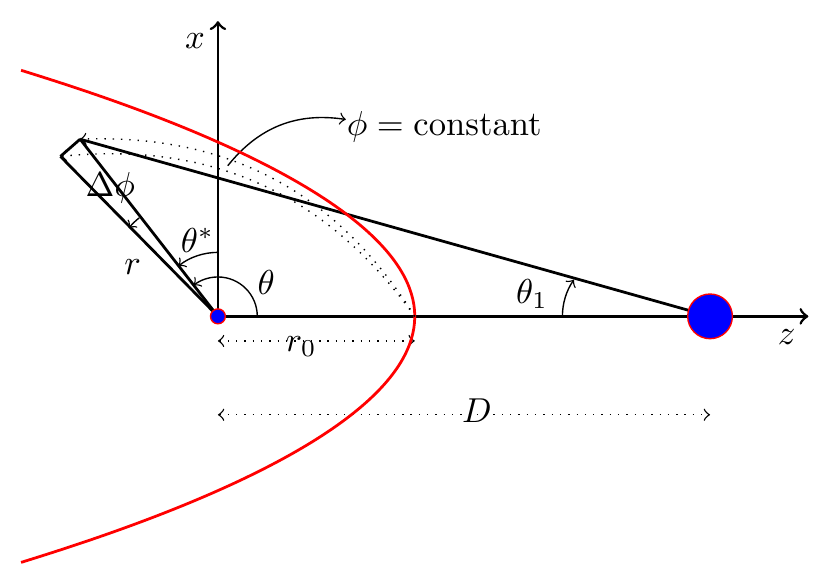}
\caption{Intrabinary shock geometry. The shocked winds are assumed to flow along $\phi$ slices of infinitesimal width $\Delta \phi$; the momentum is integrated along these slices. Here $\theta_\ast$ denotes the polar angle of the (possibly anisotropic) wind.}\label{fig:geometry}
\end{figure}

For clarity, we draw the $\beta >1$ case in Fig.\ \ref{fig:geometry} (shock wraps around the pulsar). Any chosen point on the shock is defined by the pulsar-centered angles $\theta$ from the line of centers between the stars and $\phi$ measured from the orbit normal. Such a point subtends an angle $\theta_1$ from the line of centers as measured from the companion. To connect with the anisotropic pulsar wind, we note that $\cos \theta_\ast = \sin \theta \sin \phi$. The shocked winds flow along the CD, where momentum conservation requires the total momentum flux to be the vector sum of the two wind momentum fluxes, integrated along the flow lines of constant $\phi$ \citep{2000ApJ...532..400W}.

For the pulsar wind, we have incident energy, linear- and angular- momentum fluxes as
\begin{equation}\label{eq:mass}
\mathrm{d}\dot{E}_P(\theta', \phi) = \frac{\dot{E}_{\mathrm{PSR}}}{4\pi}\sin^{2n}\theta_\ast\sin\theta'\,\mathrm{d}\theta'~,
\end{equation}
\begin{equation}
\dot{\Pi}_{Pz} = \int_0^\theta \cos\theta\,\frac{\mathrm{d}\dot{E}_P}{c}~.
\end{equation}
\begin{equation}
\dot{\Pi}_{Pr} = \int_0^\theta \sin\theta\,\frac{\mathrm{d}\dot{E}_P}{c}~,
\end{equation}
and 
\begin{equation}\label{eq:ang_mom}
\dot{J}_{P\theta} = 0~.
\end{equation}
For the case when both winds are isotropic, the corresponding equations are given in \cite{1996ApJ...469..729C}. Following their treatment, the shape of the IBS, a purely geometrical result, is given by 
\begin{eqnarray}\label{shape}
r(\theta, \phi) = D\sin\theta_1\csc(\theta+\theta_1)~,
\end{eqnarray}
where $D$ is the separation between the stars. From the momentum balance condition and fluxes (2)-(5) we can relate $\theta$ and $\theta_1$ by
\begin{eqnarray}\label{theta_1}
    \theta_1\cot\theta_1 = 1+2\beta^{-1}\csc\theta\int_0^{\theta}(1-\sin^2\phi\sin^2\theta')^n\nonumber\\
    \sin(\theta'-\theta)\sin\theta'\,\mathrm{d}\theta'~.
\end{eqnarray}
where $\beta$ is the ratio of momentum flux from the companion to that of the pulsar (Eq. \ref{eq:beta}). For $n=0$, Eq. \eqref{theta_1} recovers the standard result for two isotropic winds, whereas for $n=1$, we have
\begin{align}\label{eq:theta_1}
   \theta_1\cot\theta_1 = 1+\frac{\beta^{-1}}{8}[\theta\cot\theta(5+3\cos2\phi)-8\nonumber\\
   +(5+\cos2\theta)\sin^2\phi]~
\end{align}
and for $n=2$, we have
\begin{align}\label{eq:2theta_1}
\theta_1\cot\theta_1 = 1+\frac{\beta^{-1}}{96}\bigg[\frac{3}{2} \theta\cot\theta (5 \cos 4\phi+28 \cos 2\phi + 31)\nonumber\\
+(\cos 4\theta-14 \cos 2\theta-47)\sin^4\phi\nonumber\\
+24 (\cos 2\theta+5) \sin^2\phi-96\bigg]~.
\end{align}

\begin{figure*}
\centering
  \includegraphics[scale=0.32]{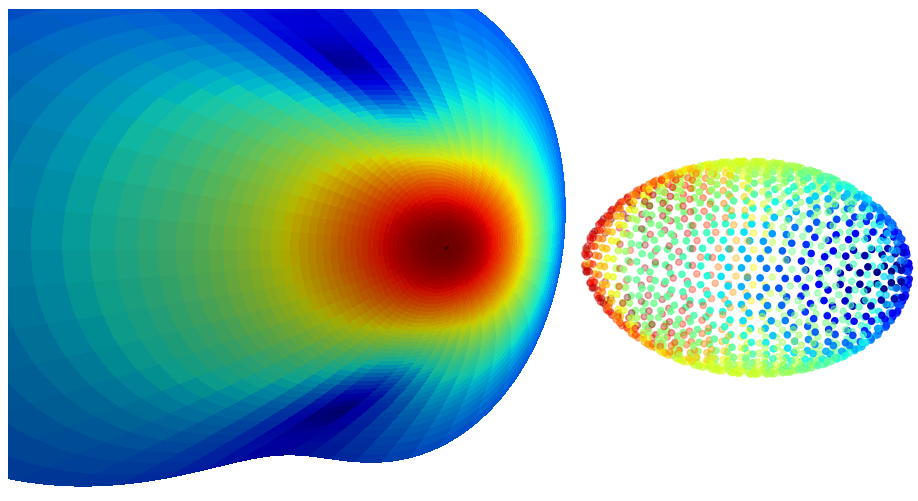}\hspace{1.6cm}
  \includegraphics[scale=0.32]{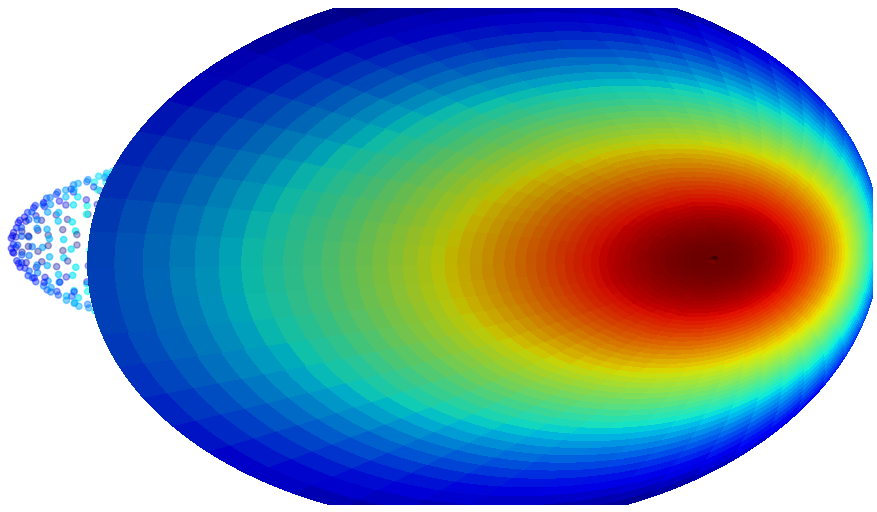}
  \caption{Shape of the CD for interaction of an anisotropic $\sin^2\theta_*$ pulsar wind with an isotropic companion wind. Left: $\beta = 10$ (RB case, shock wraps around pulsar). Right: $\beta = 0.1$ (BW case, shock wraps around companion). The shock color table maps the absorbed pulsar wind flux per unit area, while the companion's maps the direct (photon) heating power.}\label{fig:cd_shape1}
\end{figure*}

For $\beta < 1$ (the BW case), the shock wraps around the companion star so that $\theta$ at a CD location becomes the angle measured with respect to the line of centers from that star while $\theta_1$ is now the angle from the pulsar. We also re-assign $\cos \theta_\ast = \sin \theta_1 \sin \phi$ to describe the pulsar wind asymmetry. Then, the IBS geometry is described
by exchanging $\theta\leftrightarrow \theta_1$ and $\beta^{-1}\rightarrow\beta$ in Eqs. \eqref{eq:theta_1} and \eqref{eq:2theta_1}.  For $n\ne 0$, the two cases give quite different geometries. When $\beta \ll 1$ (BW), the IBS forms relatively far from the pulsar and subtends a small angle so that the shock geometry is relatively insensitive to $n$ and close to the isotropic result. In contrast, for $\beta \ga 1$, the shock surrounds the pulsar and the anisotropy increases with $n$, with the IBS increasingly flattened at the poles. The result is an `hourglass' cross-section (Fig. \ref{fig:cd_shape1}).

\section{IBS Model for synchrotron radiation} \label{sec:basics}
The pulsar outflow is a cold wind of relativistic ($\gamma_W \gg 1$) electrons and positrons of number density $n_0$ embedded in a strong magnetic field with $\sigma_w = B^2/(4\pi \gamma_w n_0 c^2)\gg 1$. Models of the pulsar magnetosphere (e.g. \citealt{2016MNRAS.457.3384T}, \citealt{2018arXiv181101767L}) suggest that the wind likely has a sector structure with the power concentrated in the equatorial plane. Recent PIC simulations \citep{2018ApJ...855...94P} suggest that the magnetization $\sigma$ may also vary with latitude and that ions may be a significant component of the outflow along some streamlines. Such variation may be important for the axially symmetric structures (equatorial tori and polar jets) seen in a number of isolated pulsar wind nebulae (PWNe). Lacking a detailed prescription for such variation, we focus here on the bulk energetics of the outflow. Beyond the light cylinder, we can assume a toroidal structure with $B\sim 1/r$ until the wind shocks. At the IBS, we expect shocks and/or reconnection to convert a fraction $\eta$ of the bulk energy to a power-law $e^\pm$ (hereafter electron) distribution with spectral index $p$ and significant pitch angle to the remnant embedded $B$. This results in synchrotron emission from the accelerated particles. To study this emission and its dependence on shock and viewing geometry, we have further extended the {\tt ICARUS IBS code} (\citealt{2012ApJ...748..115B}, \citetalias{2016ApJ...828....7R}), including cooling of and synchrotron radiation from these electrons. 

We assume that at each point on the IBS, the impinging pulsar wind converts a fraction $\eta$ of its power into an electron spectrum with energy $\gamma_{\rm min} < \gamma_e < \gamma_{\rm max}$ distributed as
\begin{eqnarray}\label{eq:injection}
N(\gamma_e)\mathrm{d}\gamma_e= N_0\gamma_e^{-p}\mathrm{d}\gamma_e~
\end{eqnarray}
and $N_0$ normalizes the sky-integrated power to $\eta {\dot E}$. The index $p$ depends on the nature of the particle acceleration. If dominated by diffusive shock acceleration (DSA), we expect $p \geq 2$, with lower values possible for oblique shocks. However, if reconnection and magnetic turbulence dominate the acceleration, we may expect a very hard powerlaw extending up to $\gamma_{max} \sim \sigma_w \gamma_w$ and indices as hard as $p \sim 1$ for $\sigma_w \gg 1$. In fact, most IBS X-ray spectra are very hard suggesting that reconnection dominates. Numerical simulations of such reconnection confirm the general trend toward harder spectrum at high magnetization \citep[e.g.][]{2014ApJ...783L..21S,2016ApJ...816L...8W} and many more recent simulations are exploring this process. Our present bulk energetics treatment of the shock radiation will not be able to address the predictions of such simulations, although it is to be hoped that, as our spectral measurements and phenomenological understanding of the IBS improve, we can compare with the particle spectral index predicted by high fidelity simulations. Note that the latitudinal variations in particle content and $\sigma$ noted above could allow the accelerated spectrum to vary across the shock; we also ignore such effects in the present treatment.

If we assume that the pulsar wind's embedded magnetic field $B$ is toroidal outside of the light cylinder $r_{LC}=c/\Omega$, then we expect
\begin{eqnarray}
B(r) = B_{0}\frac{r_0}{r}~,
\end{eqnarray}
where $B_{0} = B_{LC} (r_{LC}/r_0)=(-3I \Omega {\dot \Omega}/2cr_0^2)^{1/2}$ and $r_0$ are the magnetic field at the nose of the IBS and the shock standoff distance, respectively ($I$, $\Omega$ and ${\dot \Omega}$ are the pulsar moment of inertia, angular frequency and spin-down rate). The motion of relativistic electrons in this magnetic field leads to synchrotron radiation whose power per unit angular frequency $\omega$ for a single electron in its rest frame is given by \citep{1979rpa..book.....R}
\begin{eqnarray}
P_\omega = \frac{\sqrt{3}q^3B\sin\alpha}{mc^2}F\left(\frac{\omega}{\omega_c}\right)~,
\end{eqnarray}
where 
\begin{eqnarray}
\omega_c = \frac{3\gamma_e^2qB\sin\alpha}{2mc}~,
\end{eqnarray}
and $F(x)\equiv x\int_{x}^\infty K_{5/3}(\xi)\mathrm{d}\xi$.

Of course, synchrotron cooling depletes the electrons' energy as they travel downstream in the IBS. Thus, at a given point on the IBS, the local electron spectrum and radiated photon spectrum depend on both the freshly accelerated electrons and the upstream injection and subsequent cooling. The energy loss rate of the relativistic electron is given by \citep{1979rpa..book.....R}
\begin{equation}\label{eq:enerloss}
\gamma_e = \gamma_{e,0}(1+A\gamma_{e, 0} \tau)~,
\end{equation}
where $\tau$ is the cooling time and
\begin{equation}
A = \frac{2e^4B^2\sin^2\alpha}{3m^3c^5}~.
\end{equation}
To model the IBS electrons, we set up a grid of IBS zones and a logarithmic grid of energy bins. We then follow the injection of fresh electrons (Eq. \ref{eq:injection}) and the transfer of the electrons to lower energy bins via cooling (Eq. \ref{eq:enerloss}) as the population flows along the IBS surface with bulk motion $\Gamma$.

We expect $\Gamma$ to slowly increase along the IBS, as the hot shocked pulsar wind expands downstream. This acceleration should be slower in the RB case, where the shocked relativistic flow is confined by the massive companion wind. In this paper, we take the variation of bulk gamma along the shock as 
\begin{eqnarray}\label{ref:bulk}
\Gamma(r) = \Gamma_0\left(1+ k\frac{s}{r_0}\right)~,
\end{eqnarray}
where $\Gamma_0$ is the bulk $\Gamma$ at the nose with standoff distance $r_0$, $s$ is the arc length from the nose to a position of interest along the IBS, and $k$ is a scaling coefficient. Examining the relativistic hydrodynamic simulations of \cite{2012MNRAS.419.3426B} for the BW geometry and of \cite{2015A&A...581A..27D, 2008MNRAS.387...63B} for the RB case, we can estimate $\Gamma_0= 1.2$ and $k\sim 0.4$ (BW) and $\Gamma_0= 1.1$ and $k\sim 0.2$ (RB). This $s$ scaling is better for RB than the pulsar distance scaling used by \citetalias{2016ApJ...828....7R}, which was adequate for BW. These parameters are estimated for the bulk of the shocked wind flow along the CD. The simulations do, however, include the finite width of the shocked flow and there appears to be some variation in bulk $\Gamma$ across this width, which may be captured in future more detailed models. 

For each IBS grid zone of length $\ell_{\mathrm{grid}}$ we compute the residence time as  
\begin{equation}
\tau_{\mathrm{res}} = \frac{\ell_{\mathrm{grid}}}{c\sqrt{1-1/\Gamma_{\mathrm{grid}}^2}}~.
\end{equation}
Then electrons spend equal time (have equal weight) in this zone over the range $t=(0, \tau_{\mathrm{res}})$. We apply Eq. \eqref{eq:enerloss} to determine the re-partition of each zone's energy bins into the bins of the downstream zones. For such cooling, the number spectrum is approximately a broken power-law, with break energy $\sim 1/A$, although the detailed shape of this break and the spectrum, especially at high energies, depend on the history of the electron population as set by the upstream IBS shape and magnetic field.

The electrons flow downstream along the IBS, whose geometry is given by Eqs. \eqref{shape}, \eqref{theta_1}, and the equilibrium spectrum of electrons in each grid zone is determined by summing electrons freshly from the pulsar and cooled electrons flowing from upstream grids. To approximate the flow pattern on our numerically realized (triangularly tiled) IBS surface, we identify the two triangles adjacent to a given zone and partition the original zone's (cooled) electrons to these daughters. Such partitioning needs to capture the flow downstream and conserve the electron number at the same time. To achieve these, we start at the triangular tile at the nose and use a graph search algorithm to find its triangular neighbors, each sharing an edge with the nose tile. The parent tile is then marked as visited, and its cooled equilibrium electron population is distributed evenly to its three neighbors. The same procedure is repeated for next-generation tiles, but these three tiles and their subsequent generation will only have two unvisited neighbors. Therefore, we distribute cooled electron flux evenly among the two daughter tiles. This gives the equilibrium number spectrum of relativistic electrons in each grid zone along the IBS. 

The total power per unit frequency from the IBS synchrotron emission can finally be obtained by summing over emission from all electrons as
\begin{eqnarray}
L_\omega = \int \mathrm{d}\gamma_e~\,N(\gamma_e)P_\omega(\gamma_e)~.
\end{eqnarray}
The above expression is valid in the rest frame of electrons. In order to obtain observed power, one needs to account for Doppler boosting and beaming. Doing so, we calculate observed $L_\omega$ from each triangular grid at a given sky angle, and sum over the entire CD to obtain the flux at each sky position. This can be represented in discretized form by
\begin{eqnarray}\label{eq:main_flux}
L_\omega = \sum_j\frac{1}{\Gamma_j^3\left(1-\sqrt{1-\frac{1}{\Gamma_j^2}}\cos\theta_j\right)^3}\times\nonumber\\
\int \mathrm{d}\gamma_e~\,N_j(\gamma_e)P_{\omega_j}(\gamma_e)~,
\end{eqnarray}
where index $j$ represents each triangular grid zone of the CD, $\theta_j$ the angle between flow direction on a triangular grid and the sky direction, and $\omega_j$ is plasma frame photon frequency which is Doppler shifted to the observed frequency $\omega$ by
\begin{eqnarray}
\omega = \frac{\omega_j}{\Gamma_j\left(1-\sqrt{1-\frac{1}{\Gamma_j^2}}\cos\theta_j\right)}~.
\end{eqnarray}
The first factor inside the sum in Eq. \eqref{eq:main_flux} follows from the Lorentz invariance of phase space density of photons during transformation between plasma frame and observer's frame.

Finally, Eq. \eqref{eq:main_flux} can be used to produce energy-resolved sky-maps of the synchrotron emission from the IBS. In turn, for a given viewing angle at binary inclination $i$, we can extract the observable phase-resolved spectrum (or energy-dependent light curves).

\begin{figure*}
\centering
\includegraphics[scale=0.44]{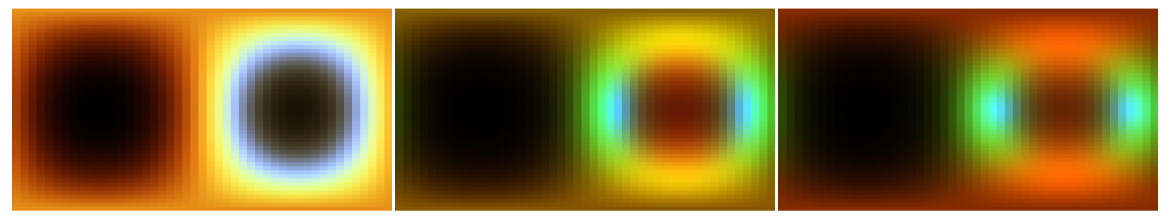}
\caption{Sky map of emission from the IBS. Left is for isotropic, center for $\sin^2\theta_*$ and right for $\sin^4\theta_*$ distribution of the energy flux from the pulsar. Red corresponds to emission from near the nose of the CD, i.e. $\theta < 30^\circ$, green from $30^\circ<\theta<60^\circ$ and blue from $\theta>60^\circ$. In all cases, the peak emission comes mainly from downstream CD regions, while off-pulse emission comes mainly from the nose.}\label{fig:sky_map}
\end{figure*}

\section{Properties of IBS synchrotron emission}\label{sec:3}

Fig. \ref{fig:sky_map} shows sky maps of emission from the IBS for a RB geometry for isotropic, $\sin^2\theta_\ast$ and $\sin^4\theta_\ast$ winds. The general geometry is a ring of emission surrounding pulsar inferior conjunction $\phi_B=0.75$. As the Earth line-of-sight cuts through this ring near the orbital plane one sees a light curve with two caustic peaks bracketing a fainter `bridge' region. A more polar view can have a single peak or no peaks. For a finite velocity companion wind, the peaks can be delayed and asymmetric (\citetalias{2016ApJ...828....7R}). In general, the brightest emission comes from downstream on the CD, where the bulk $\Gamma$ is large, increasing beaming and Doppler boosting, and the IBS surface is flatter, leading to brighter caustic peaks in the light curves. The colors show that the regions further downstream are most important at the inner edges of the peaks. Since these downstream regions have experienced the most cooling, this can lead to (subtle) energy dependence in the synchrotron light curves. We see that the peak-bridge flux ratio is largest in the isotropic case. This is because for the anisotropic case, the IBS downstream from the nose is flatter and more tangential to the pulsar wind so that the electron power away from the nose is smaller. Thus, the broadly distributed emission from low $\Gamma$ nose regions is more important; such zones radiate to more of the sky, adding flux to the bridge.

\begin{figure}
\centering
\includegraphics[scale=0.41]{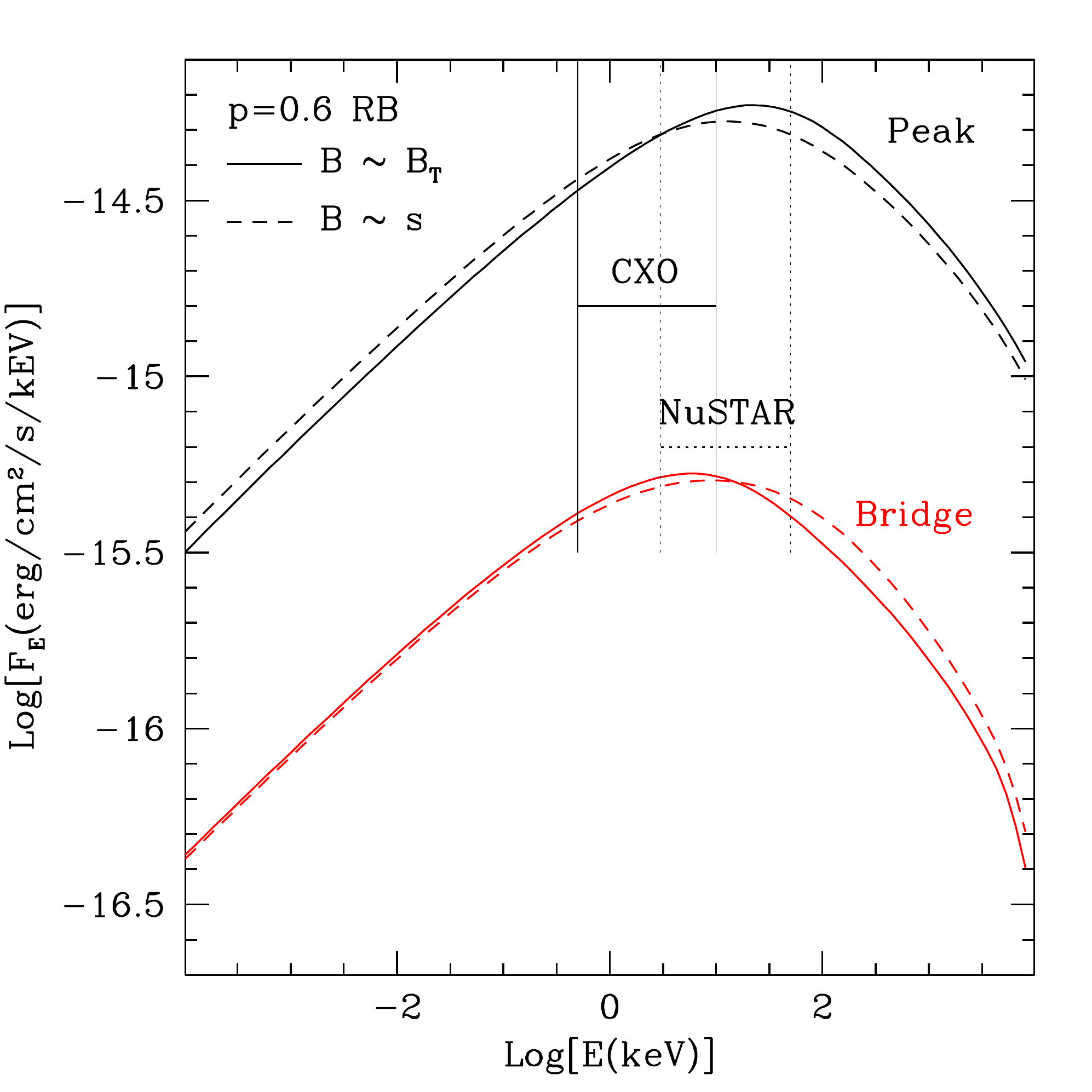}
\caption{Spectrum of emission from the peak and the bridge for a RB model, assuming electrons with $p=0.6$ and nose field $B_0=100$\,G. This $B$ either follows a toroidal pattern (solid) or increases along the CD (dashed). The cooling break at the peak is higher than at the bridge, as can be probed with X-ray measurements. This difference is smaller for $B$ increasing with $s$.}\label{fig:spectra}
\end{figure}

Since synchrotron electrons are cooled as they flow downstream, there is a break in the spectrum. Interestingly, for typical parameters the break occurs in the X-ray band. The peak flux includes emission from farther along the CD, with larger Doppler factor. Thus, in the observer's frame, the peak emission has a higher break energy, as shown in Fig. \ref{fig:spectra}. However, the separation depends on the cooling and hence the IBS B field. While the solid curves in Fig. \ref{fig:spectra} are for a field swept up from the pulsar wind, one might also imagine a shock-generated field increasing along the IBS. For this case (dashed lines), the cooling rate grows so that bridge and peak are from very similar regions with similar break energies. For typical parameters, both peak and bridge breaks are accessible if one includes hard X-rays. Even in the soft X-ray band, this difference may be discerned by small changes in the effective photon index.  

\begin{figure}
\centering
\includegraphics[scale=0.41]{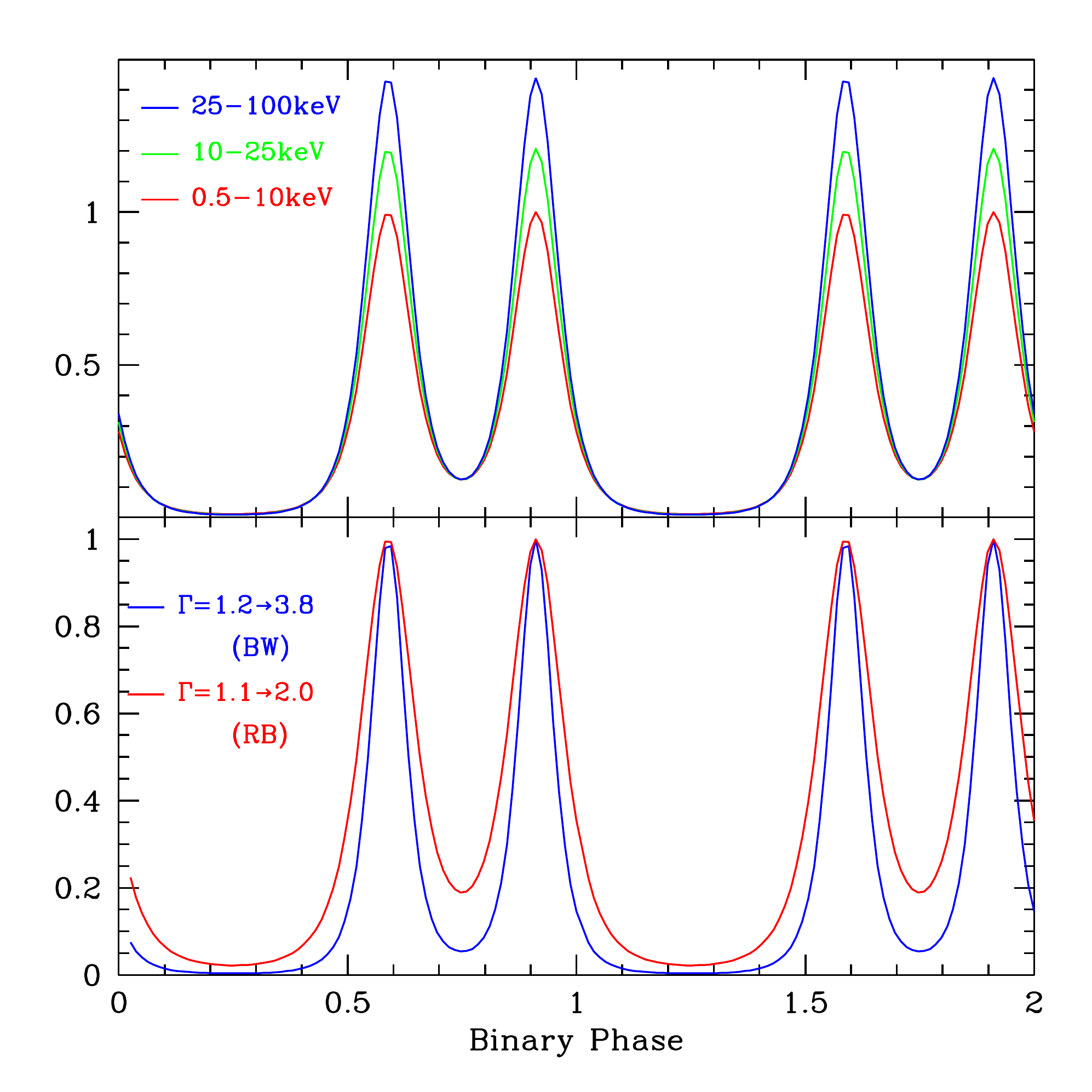}
\caption{Top: Light curves for different energy bands of 0.5 - 10 keV, 10 - 25 keV and 25 - 100 keV for the toroidal $B$ model of Fig. \ref{fig:spectra}. The peak to bridge contrast is larger at high energies. Bottom: lightcurves for BW-type and RB-type bulk gamma growth. }\label{fig:light_curves}
\end{figure}

These spectral changes induce energy dependence in the light curves (Fig. \ref{fig:light_curves}, top). These light curves are also sensitive to the growth of the bulk Doppler factor along the IBS. For black widows, the acceleration of \cite{2012MNRAS.419.3426B} generates narrow peaks and faint bridge, while the slower $\Gamma$ growth expected for RB leaves larger bridge emission. The finite width of the shock also segregates cooled electrons from fresh injection; this can lead to additional energy dependence, which we describe in a future communication.

\section{Application of the model}
\subsection{PSR J1959+2048}
PSR J1959+2048 = PSR B1957+20 (hereafter J1959) is a BW system with a $P=1.6$ ms, $\dot{E} = 1.6\times 10^{35}$ erg s$^{-1}$ millisecond pulsar in a $P_b=9.1$\,hr orbit with a $\sim 0.025 M_\odot$ companion. Modeling the optical and X-ray orbital variations can provide important information on the system masses and the wind physics. 

\begin{figure*}
\centering
\includegraphics[scale=0.4]{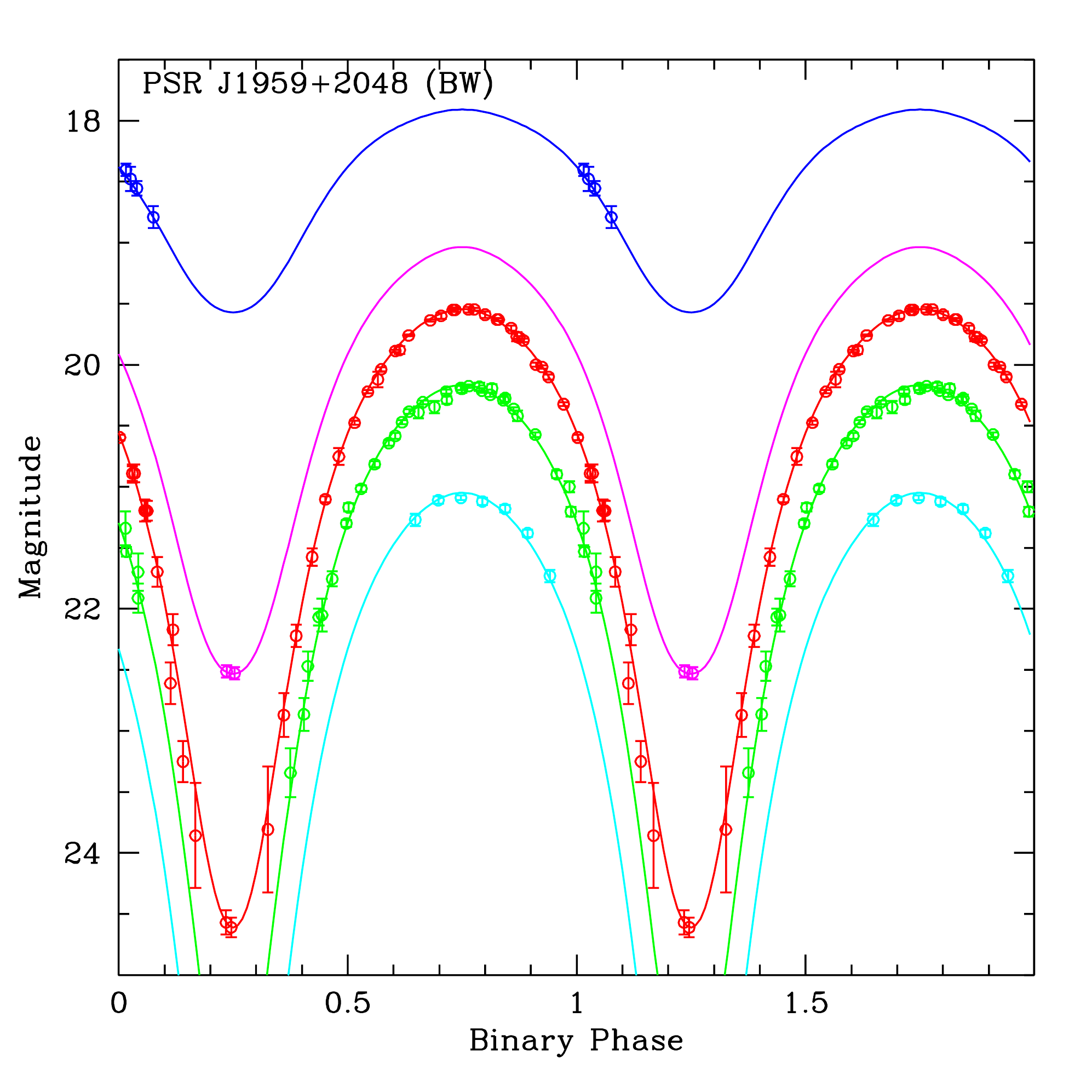}\hspace{0.2cm}
\includegraphics[scale=0.4]{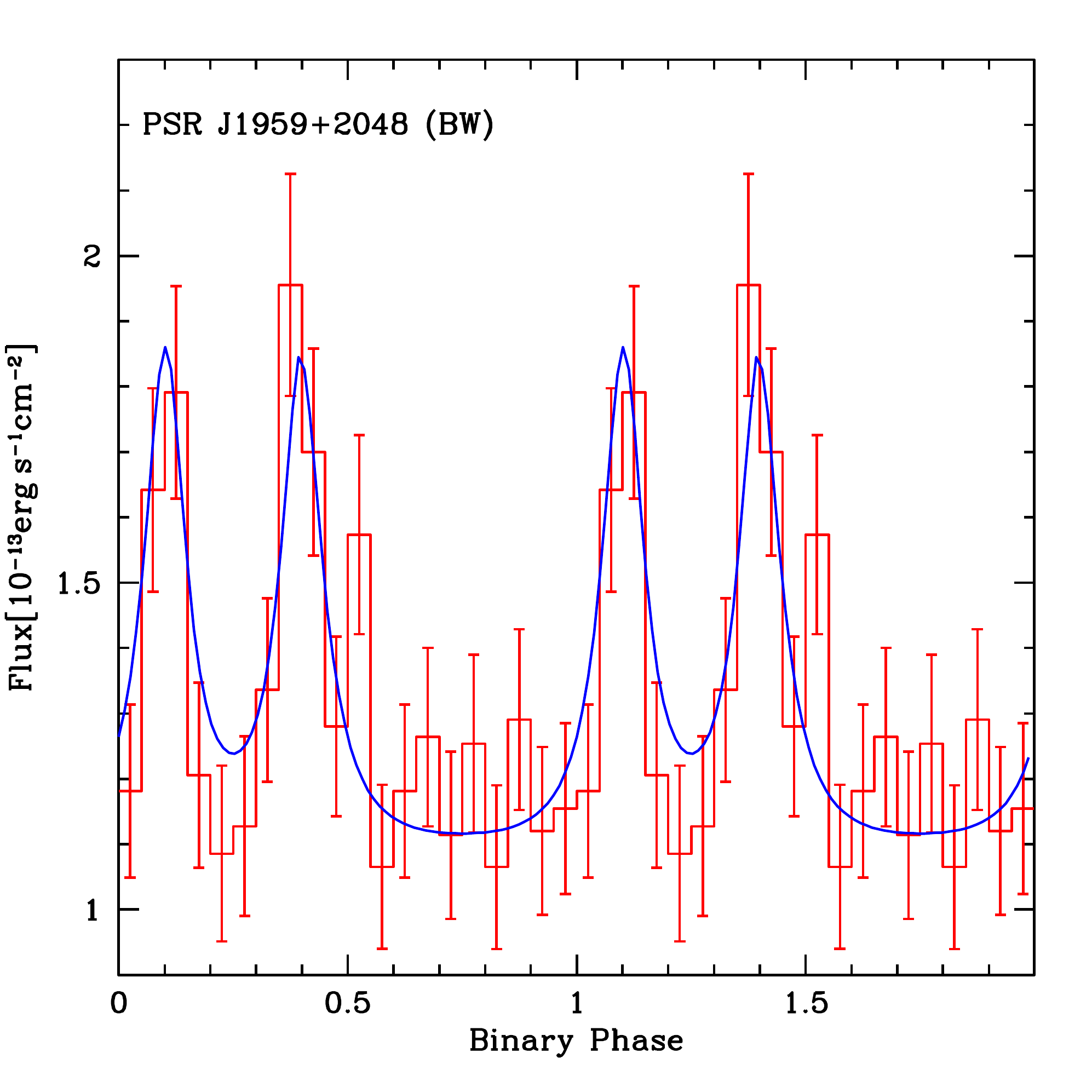}
\caption{Left: BVRIKs Optical lightcurves of J1959, compared with the direct heating model from the best-fit parameters. Right: Combined {\it CXO} and {\it XMM} lightcurve of J1959 for energy range of $0.5-7$ keV, and IBS model from the best-fit parameters.}\label{fig:J1959lc}
\end{figure*}

For the optical data, we used the BVRIK$_s$ magnitudes from \citet{2007MNRAS.379.1117R} and assumed an extinction $A_V=0.82$. As shown in Fig. \ref{fig:J1959lc}, the direct heating model provides a reasonable fit to the data, and results in an inclination $i\sim 63.9^\circ \pm 1.2^\circ$, a distance $2.27\pm 0.02$ kpc. This angle is consistent with the findings of \citet{2011ApJ...728...95V} and for their measured radial velocity implies a neutron star mass $\sim 2.38M_\odot$. The distance is somewhat smaller than the 3.3 kpc estimated in SR17; it is now consistent with the largest DM estimates, but the required direct heating, a sky-integrated $5 \times 10^{34} {\rm erg\, s^{-1}}$, is still substantially larger than inferred from the observed (mostly GeV) photon flux,
which gives an isotropic equivalent luminosity $1.05 \times 10^{34} {\rm erg\, s^{-1}}$. It is also a large fraction of the Shklovskii-corrected spindown luminosity of $7.8 \times 10^{34} I_{45} {\rm erg\, s^{-1}}$; we thus conclude that the companion sees more pulsar flux than directed at Earth and, likely, the moment of inertia is larger than the default $10^{45} {\rm g\, cm^2}$.

Archival X-ray observations of this object include 212 ks of CXO ACIS-S exposure (Obs ID 9088, 1911) and 31.5 ks of XMM exposure (Obs ID 0204910201). Unfortunately, the XMM observation was made in PN timing mode (with high background) so that only the MOS data were useful for this orbital study. Also, the observation did not cover the `Br' phase between the peaks, severely limiting the effective XMM exposure. Nevertheless, we employed these data to extract a combined light curve and spectra, using {\tt HEAsoft}, {\tt CIAO} and {\tt SAS}. Spectra were measured in four orbital phase bins (Table \ref{tab:spec}) with the analysis showing hard power law emission with (low significance) variation between phases. As described in Sec. \ref{sec:3}, this may be understood if the cooling breaks falls within the X-ray range. 

The light-curve shows a nice double-peaked structure centered at $\phi_B = 0.25$ (Fig. \ref{fig:J1959lc}). To model the IBS emission, we estimate the magnetic field strength by assuming a toroidal field outside of the light-cylinder and $I\sim 10^{45}$ \,g\,cm$^2$ together with the orbital parameters, which give $B_0\sim 20$ G. Fitting this lightcurve with our model (for a $\sin^2\theta_*$ wind) gives $i\sim 75.8^\circ\pm 5.9^\circ$ (see Table \ref{tab:fit}), which is higher than that inferred for the direct heating model (e.g. \citealt{2011ApJ...728...95V}, \citealt{2017ApJ...845...42S}). With such inclination, the mass of neutron star would be $\sim 1.86M_\odot$, a value much smaller than inferred through optical fit. In a combined fit, the optical points with small errors dominate, giving an inclination of $64.12^\circ \pm 1.35^\circ$, close to the optical-only fit. The x-ray and combined fit suggest $\beta\sim 0.12$, signifying a very strong pulsar wind. 

We do note some X-ray peak asymmetry which (in higher S/N data) may be used to constrain sweepback due to the finite speed of the companion wind via the parameter $f_v$ in RS16. This is interesting as, together with $\beta$, one can infer the mass loss rate to see if complete companion evaporation is expected.

\begin{deluxetable}{cllll}

\tabletypesize{\footnotesize}
\tablewidth{0pt}

 \tablecaption{ Spectral fit results \label{tab:spec}}

 \tablehead{
 \colhead{Phase} & \colhead{J1959 $\phi$} & \colhead{J1959 $\Gamma$} & \colhead{J2339 $\phi$} & \colhead{J2339 $\Gamma$}
 }

 \startdata 
 Off & 0.55-0.05 & $1.90\pm 0.19$ & 0.05-0.45 & $1.19\pm 0.11$\\
 P1 & 0.05-0.185 & $0.72\pm 0.44$ & 0.45-0.65 & $0.96\pm 0.11$ \\
 Br & 0.185-0.32 & $1.37\pm 0.89$ & 0.65-0.85 & $0.80\pm 0.13$\\
 P2 & 0.32-0.55  & $1.08\pm 0.35$ & 0.85-0.05 & $0.77\pm 0.12$
 \enddata
\vspace{0.2cm} 
\tablecomments{$\Gamma$ represents the IBS PL index on top of a fixed thermal (NS surface) plus power-law (NS magnetosphere, large scale PWN) background from the Off interval.} 
\end{deluxetable}

\begin{deluxetable*}{lccccc}

\tabletypesize{\footnotesize}
\tablewidth{0pt}

 \tablecaption{ Model fit results \label{tab:fit}}

 \tablehead{
 \colhead{Parameters} & \colhead{$J1959_X$} & \colhead{$J1959_O$} & \colhead{$J1959_{O+X}$} & \colhead{$J2339_X$}}
 
 \startdata 
 $\dot{E}_{\mathrm{IBS}}(10^{34}\mathrm{erg}/\mathrm{s)}$ & $8.67 \pm 1.40$  & $4.76\pm 0.10$  & $6.03\pm 0.13$ & $4.70 \pm 0.30$\\
  $i \mathrm{ (}^\circ)$ & $75.8\pm 5.9$  & $63.9\pm 1.2$ & $64.12 \pm 1.35$ & $54.05 \pm 6.38$\\
 $\beta$ & $0.118\pm 0.015$  & -  &  $0.127 \pm 0.018$ & $5.19 \pm 0.23$\\
 $T_N$ (K) & - & $2743\pm 29$ & $2748\pm 33$  & -\\
 $d_{kpc}$ & 2.27  & $2.27\pm 0.02$  & $2.27\pm 0.02$ & 1.25\\
 $f_1$ & -  & $0.886 \pm 0.007$  & $0.887\pm 0.008$ & -\\
 $M_{\mathrm{NS}}(M_\odot)$ & 1.86  & 2.37  & 2.33 & 2.16\\
  $\chi^2/\nu$ & 1.12  & 1.54  & 1.49 & 2.57
 \enddata
 \tablecomments{$M_{\mathrm{NS}}$ from fit parameters and literature companion radial velocity.} 
\end{deluxetable*}
\subsection{PSR J2339-0533}

PSR J2339$-$0533 is a $P_s = 2.9$\,ms, ${\dot E}= 1.7 \times 10^{34} {\rm erg\, s^{-1}}$ `redback' MSP in a 4.6\,h orbit about a $\sim 0.3 M_\odot$ companion. The companion wind seems particularly strong, with nearly continuous radio eclipses. This wind may also be associated  with the large orbital period instabilities \citep{2015ApJ...807...18P}. With a modest ${\dot E}$ and a strong companion wind, it is not surprising that the observed X-ray IBS peaks bracket $\phi_B=0.75$ so that $\beta>1$ and the shock wraps around the pulsar. What is more surprising is the remarkably hard X-ray spectral indices with several components exhibiting $\Gamma$ significantly less than 1. We infer that reconnection dominates the electron energization and, with the RB geometry, significant IBS flux comes from regions at high pulsar latitude (i.e small $\theta_\ast$). These regions, away from the equatorial plane, may have a large $\sigma_w$ and small particle content {\citep[see][]{2018ApJ...855...94P}}. We can speculate that at these IBS positions, the X-ray spectrum is affected by large $\gamma_{\rm max} \approx \sigma_W \gamma_w$ with the local Maxwellian peak leading to a particularly hard spectrum. It seems unlikely that $p < 1$ can persist over many energy decades. This hard injection index is used to illustrate cut-off sensitivity in Fig. \ref{fig:spectra}.

\begin{figure}
    \centering
    \includegraphics[scale=0.4]{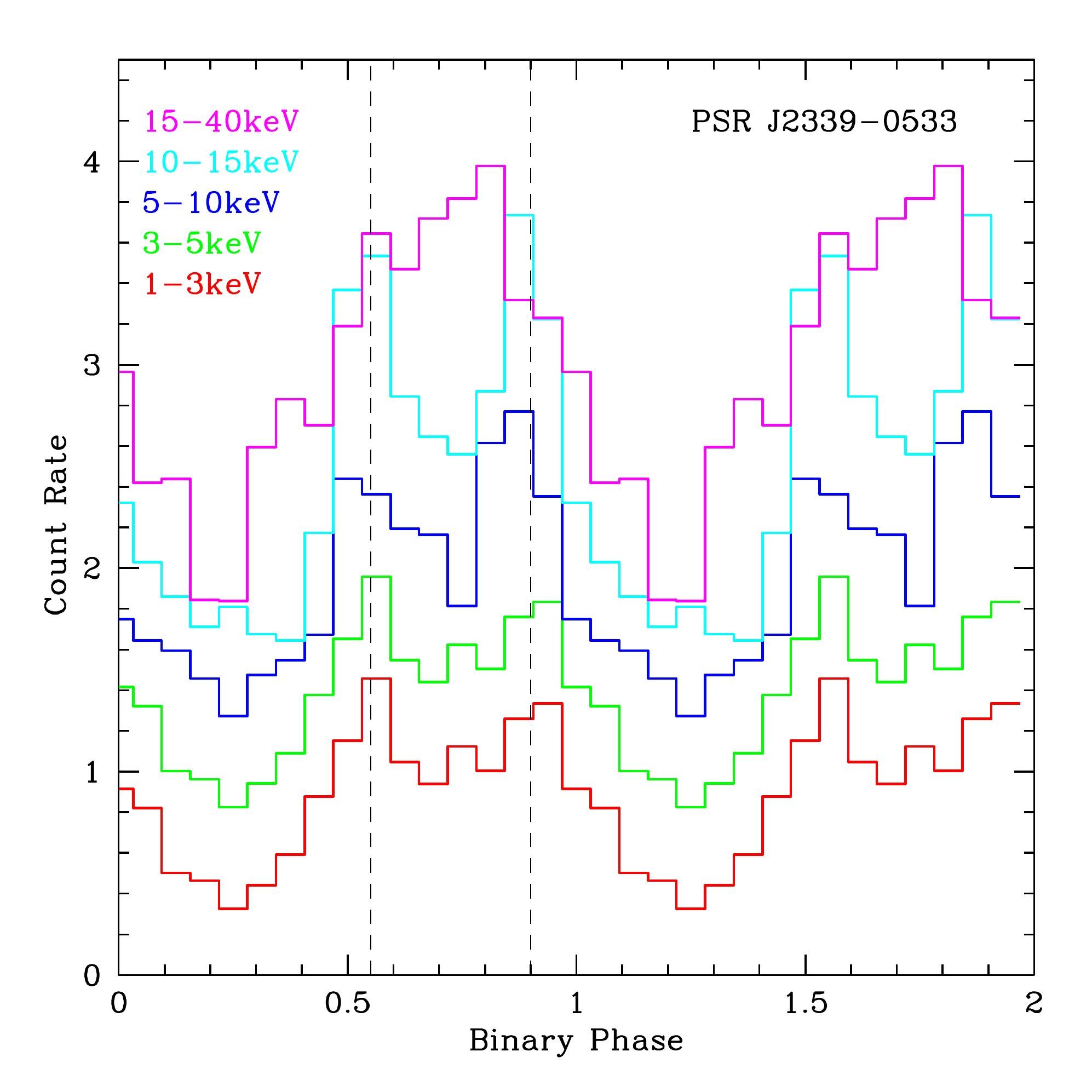}
    \caption{Combined {\it CXO}, {\it XMM} and NuSTAR light curves for RB J2339. Higher energy curves are offset in y.}
    \label{fig:j2339_xrayE}
\end{figure}

To study the X-ray IBS emission, we use archival data from {\it Chandra} (20\,ks ObsID 11791), {\it Swift} (49.4\,ks cumulative), {\it XMM-Newton} (182\,ks ObsID  721130101, 790800101), {\it Suzaku} (104\,ks ObsID 406007010) and {\it NuSTAR} (163\,ks ObsID 30202020002).  We note that the second {\it XMM-Newton} observation (rev. 3121) is contemporaneous with {\it NuSTAR} one for simultaneous broadband X-ray coverage. Data were analyzed with the appropriate response functions and fitted simultaneously using {\tt XSPEC}. For this relatively bright system, we have sufficient counts to examine the orbital light curve in several X-ray bands. To do this, we made effective area/exposure-weighted count light curves, which are displayed with higher energy bands offset in rate in Fig. \ref{fig:j2339_xrayE}. The double peak bracketing $\phi_B=0.75$  with separation $\Delta \phi_B=0.35$ is quite prominent, with a strong bridge, especially at low energy. There may be a slight trend toward decreasing $\Delta \phi_B$ with energy. These behaviors are as expected from \S4.

The orbital modulation persists to $>$40\,keV (chance probability for a constant light curve being $p\approx 10^{-4}$), but at $E> 15$\,keV the double peaks disappear and a new, single or tight double component arises centered at $\phi_B=0.75$. While the statistics are insufficient for a strong conclusion, the double peak cut-off appears steeper than a cooling break, suggesting that we may be probing their electron $\gamma_{max}$ in the hard X-ray band. Similarly, the high energy `bridge' peak appears abruptly suggesting a very hard spectrum for this component.

Fitting simple absorbed power-laws to the combined data sets confirms a very hard $\Gamma \sim 1$ across all phases, with an indication that the off-pulse index is softer than that of peak (see Table \ref{tab:spec}). 

\begin{figure}
    \centering
    \includegraphics[scale=0.4]{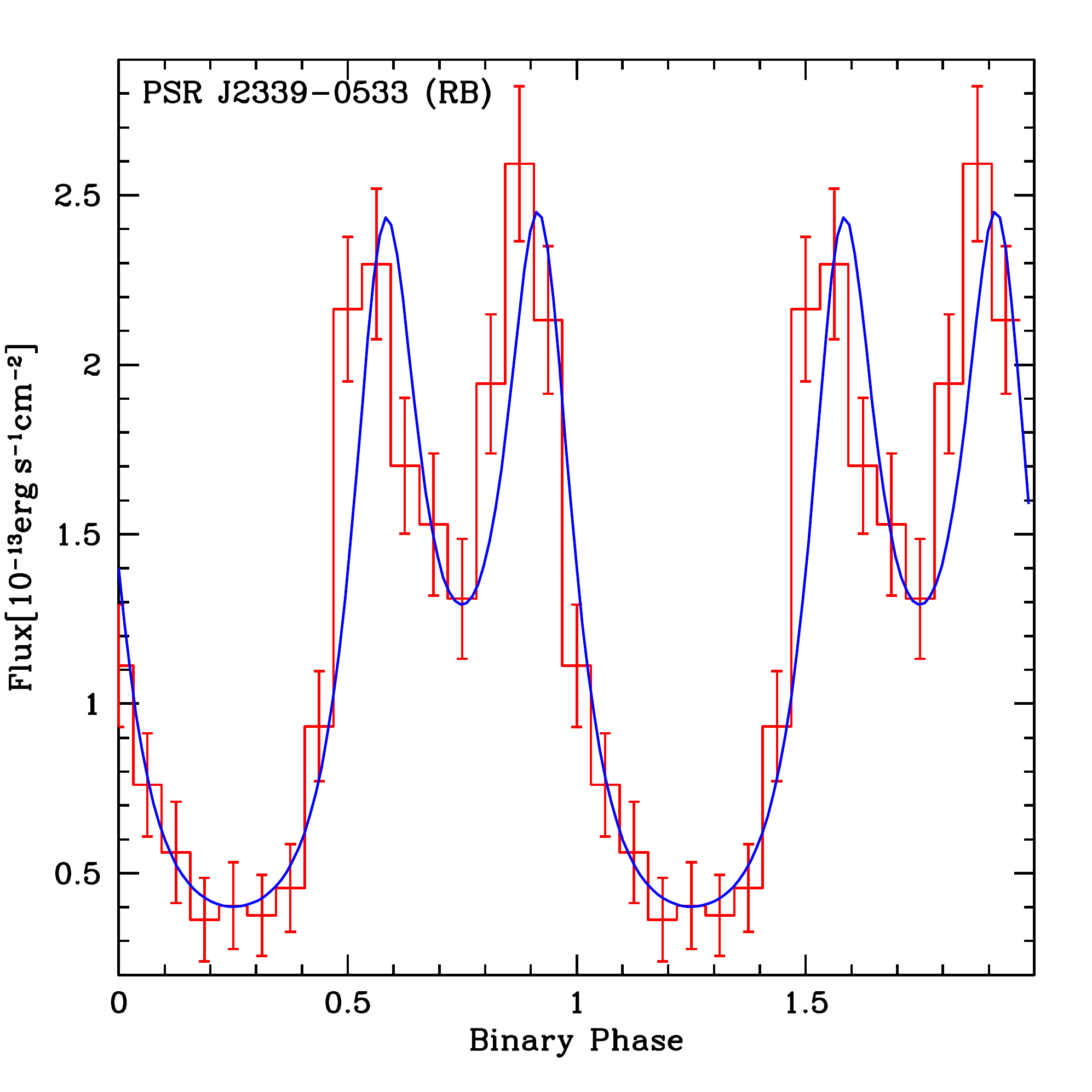}
    \caption{Combined {\it CXO}, {\it XMM} and NuSTAR $3-15$ keV light curve for J2339, compared with the best-fit IBS model.}
    \label{fig:j2339_xray}
\end{figure}

For modeling, we concentrate on the well-defined double-peaked $3-15$ keV light curve in Fig. \ref{fig:j2339_xray}. The magnetic field $B_0\sim 30$ G used in the IBS model is estimated by using typical pulsar $I\sim 10^{45}$\,gm\,cm$^2$ together with the orbital parameters. The model fitting gives $i \sim 54.1^\circ \pm 6.4^\circ$. This low inclination is consistent with previous findings of \cite{2011ApJ...743L..26R} and \cite{2015ApJ...802...84Y}, and yields an inferred neutron star mass of $\sim 2.16M_\odot$.

Note the the 3-15\,keV peaks significantly lead the model peaks. We have also collected high S/N multi-band optical light curves for J2339, and these, too show a highly significant shift to earlier phase. This behavior is commonly seen in high quality RB light curves. As described in SR17, it may be a consequence of companion magnetic fields which can perturb the wind flow and duct IBS particles to the surface for anisotropic heating. Thus, we do not attempt a joint optical/X-ray fit here, deferring it to a future publication.

\section{Conclusions}

We have examined the intrabinary shock emission from colliding winds in `spider'-type companion-evaporating pulsars, finding interesting sensitivity to the anisotropy of the pulsar wind. This leads to important differences in the contact discontinuity shape between the `redback' (RB) case where the companion wind's momentum dominates and the pulsar-dominated `black widow' (BW) case. For both, the shock produces a very hard X-ray spectrum, implying that reconnection provides the principal energization of $e^\pm$ in the shock of the high $\sigma$ pulsar wind. 

By following the accumulation and cooling of these electrons along the IBS, we see that the resulting synchrotron emission is a useful probe of the shock structure. The caustic peaks from the mildly relativistic flow along the CD are sensitive to the adiabatic acceleration of the shocked wind and the cooling break (and possibly the upper energy cut-off) are sensitive to the magnetic field and the flow speed.

We have applied these models (for an assumed ${\rm sin^2\theta_\ast}$ wind) to a BW (PSR J1959$+$2048) and a RB (PSR J2339$-$0533) with X-ray light curves showing strong IBS peaks. The data fits give us constraints on the wind parameters (e.g $\beta$) and the viewing geometry. These estimates can, at least for the BW cases which seem to be dominated by direct heating, be compared with values inferred by fits to the companion optical data. For J1959, the X-rays prefer somewhat higher inclination (and lower neutron star mass), but higher S/N X-ray data will be needed for a very constraining comparison. For the RB J2339, direct heating does not appear to explain the optical light curves and so models including magnetic ducting or other non-axisymmetric heat distribution are needed. However, in both cases, IBS fits provide information on the pulsar and companion winds beyond that obtainable from optical data alone.

\begin{acknowledgements}
This work was supported in part by NASA grant 80NSSC17K0024.
\end{acknowledgements}
\software{ICARUS (\citealt{2012ApJ...748..115B}, \citetalias{2016ApJ...828....7R}), HEAsoft, CIAO, SAS}
\bibliographystyle{aasjournal}
\bibliography{mainpaper}
\end{document}